%%%%%%%%%%%%%%%%%%%%%%%%%%%%%%%%%%%%%%%%%%%%%%%%%%%%%%%%%%%%%%%%%%%%%%%%%%%%%%
%%%%% This Tex file inputs phyzzx.       				  %%%%
%%%%%%%%%%%%%%%%%%%%%%%%%%%%%%%%%%%%%%%%%%%%%%%%%%%%%%%%%%%%%%%%%%%%%%%%%%%%%%
\input phyzzx
\def\dg{\dagger} 

%Start titlepage
\Pubnum={UUITP-07/96\cr}
\date={18 March 1996}
\titlepage
\title{Quasihole wavefunctions for the Calogero model}
\bigskip
\author {Alexios P. Polychronakos\footnote\dagger
{poly@calypso.teorfys.uu.se}}
\address{Centre for Advanced Study, Norwegian Academy of Science and Letters,
0205 Oslo, Norway}
\andaddress{Theoretical Physics Dept., Uppsala University\break
S-751 08 Uppsala, Sweden}
\bigskip
\abstract{
The one-quasihole wavefunctions and their norms are derived for
the system of particles on the line with inverse-square interactions
and harmonic confining potential.
}

\vfill
\endpage

\def\NP{{\it Nucl. Phys.\ }}
\def\PL{{\it Phys. Lett.\ }}
\def\PRD{{\it Phys. Rev. D\ }}
\def\PR{{\it Phys. Rev. \ }}
\def\PRL{{\it Phys. Rev. Lett.\ }}

\def\IJMP{{\it Int. Jour. Mod. Phys.\ }}

\REF\C{F.\ Calogero, {\it Jour. of Math. Phys.}
{\bf10}, 2191 and 2197 (1969); {\bf 12}, 419 (1971).}
\REF\S{B.\ Sutherland, \PR {\bf A4}, 2019 (1971); \PR {\bf A5}, 1372 (1972);
\PRL {\bf 34}, 1083 (1975).}
\REF\M{J.\ Moser, ``Dynamical Systems, Theory and Applications,"
Lecture Notes in Physics {\bf 38}, Springer-Verlag, New York 1975.}
\REF\LM{J.M.\ Leinaas and J.\ Myrheim, \PR {\bf B37}, 9286 (1989).}
\REF\PA{A.P.\ Polychronakos, \NP {\bf B324}, 597 (1989).}
\REF\H{F.D.M.\ Haldane, \PRL {\bf 67} (1991) 937.}
\REF\I{S.B.\ Isakov, \IJMP {\bf A9} (1994) 2563.}
\REF\MPA{J.A.\ Minahan and A.P.\ Polychronakos, \PR {\bf B50} (1994) 4236.}
\REF\LPS{F.\ Lesage, V.\ Pasquier and D.\ Serban, \NP {\bf B435}, 585 (1995).}
\REF\HA{Z.N.C.\ Ha, \PRL {\bf 73}, 1574 (1994); ERRATUM, \PRL {\bf 74},
620 (1995).}
\REF\STA{R.\ P.\ Stanley, {\it Adv.\ Math.} {\bf 77}, 76 (1989).}
\REF\F{ P.\ J.\ Forrester, \PL {\bf A179}, 127 (1993).}
\REF\PB{A.P.\ Polychronakos, \PRL {\bf 69}, 703 (1992).}
\REF\BHV{L.\ Brink, T.H.\ Hansson and M.A.\ Vasiliev, \PL {\bf B286},
109 (1992).}
\REF\N{Y.\ Nambu, in ``Gotsman, E.\ (Ed.), Tauber, G.\ (Ed.): From SU(3) 
to Gravity," pp.45-52 and \PRD {\bf 26} (1992) 2875 and references therein.}
\REF\STO{M.\ Stone, \PR {\bf B42} (1990) 399.}
\REF\J{A.\ Jevicki, \NP {\bf B376} (1992) 75.}
\REF\MPB{J.A.\ Minahan and A.P.\ Polychronakos, \PL {\bf B326}, 288 (1994).}
\REF\G{M.\ Gaudin, Saclay preprint SPhT-92-158.}

The Calogero-Sutherland-Moser class of models of interacting particles in one 
dimension [\C-\M] has received a great deal of attention, mostly due to its 
interesting mathematical properties and connection to fractional statistics
[\LM-\I]. 

It is of interest
to have explicit energy wavefunctions for these systems, since they are
needed to calculate correlation functions. So far, it is mostly the
periodic (Sutherland) model that has been used for the purpose of such 
calculations [\MPA-\HA], and its wavefunctions (termed Jack polynomials
in the mathematical literature [\STA]) have been extensively studied [\F]. 
The original harmonic (Calogero) system, on the other hand, has been 
rather neglected in this respect, due, mainly, to its translation 
non-invariance. Its wavefunctions are, likewise, rather obscure.
Although in principle they can be obtained either with the original
diagonalization method [\C] or with the operator method [\PB,\BHV], their
general explicit form is unknown. The original wavefunctions found by 
Calogero are, to this day, the only explicitly known ones. Nevertheless, 
this system has the advantage of possessing ladder operators, 
and would thus be more suited to an algebraic approach.

In this note, we present a set of wavefunctions of interest in the
many-body properties of this system, namely the one-hole wavefunctions,
along with their norms. Generically, hole states are simpler than 
particle states in these systems.
For instance, the one-hole wavefunctions of the Sutherland model are
identical to the free fermion ones, upon division by the Vandermonde
determinant in the relevant power. (Cf. also the Laughlin states.)
As we will see, the Calogero holes are not so simple, but they are
still amenable to a complete analysis.

The model of consideration is described by the hamiltonian
$$
H = \half \sum_{i=1}^N p_i^2 + \sum_{i>j} {l(l-1) \over ( x_i - x_j )^2}
+ \half \sum_{i=i}^N x_i^2
\eqn\H$$
where we chose units such that the particle masses and oscillator
frequency be one. Particle statistics are irrelevant, due to the
impenetrability of the mutual potential, and we shall consider symmetric
wavefunctions. The dynamics are determined by the coupling $l$, with
$l=0,1$ corresponding to noninteracting bosons or fermions respectively.
For $l=1$ we have the usual picture of the Fermi sea with particle and
hole excitations. By following these states as they evolve for $l \neq
1$, we are led to the notion of (quasi)particle and (quasi)holes.

For noninteracting particles, the wavefunction of a particle at 
energy $n$ in the coherent state representation is simply $z_i^n$, 
and the action of the oscillator creation operator $a_i^\dg$
is multiplication by $z_i$. The wavefunctions are therefore identical in
form to the momentum eigenstates of free particles on the circle (for
which $z_i =\exp(ix_i )$). In the case of fermions, thus,
the known connection of free fermions on the circle to representations
and characters of $U(N)$ [\N-\MPB] carries over to the $l=1$ harmonic system 
as well. Each excited state of the fermion system can be mapped to a $U(N)$
Young tableau, and thus to an irrep $R$ of $U(N)$. This is done by mapping
the particle excitations to successive rows of the tableau. 
Equivalently, each hole excitation can be mapped
to a column of the tableau. Since a hole excitation at energy $n$ is
the same as $n$ particles excited by one unit, the two pictures are
dual descriptions of the same quantum state. A general excited state
in the fermion case can be obtained as
$$
|R,f> = \chi_R ( a_i^\dg ) |0,f> 
\eqn\Rf$$
where $|0,f>$ is the fermionic $N$-body ground state and $\chi_R$ 
is the character of the representation $R$ expressed in terms of the
operator matrix $diag(a_1^\dg, \cdots a_N^\dg )$. 
The one-hole state, in particular, corresponds to a tableau 
with a single column, that is, the $n$-fold fully antisymmetric 
representation of $U(N)$. It is then expressed as
$$
|n,f> = \chi_n ( a_i^\dg ) |0,f> = \sum_{\rm distinct} 
a_{i_1}^\dg \cdots a_{i_n}^\dg |0,f>
\eqn\hf$$
where the sum is over all combinations of distinct indices.

The corresponding states in the bosonic case, obtained by ``collapsing"
each fermionic state in a way that successive particle distances in the
energy spectrum are reduced by one unit, can similarly be expressed as
$$
|R,b> = \sum_{\rm distinct} (a_{i_1}^\dg)^{n_1} \cdots 
(a_{i_N}^\dg)^{n_N} |0,b>
\eqn\Rb$$
where $n_1 , \dots n_N$ are the lengths of the rows of the Young tableau.
The ``one-hole" state thus becomes simply a state with $n$ particles
in the first level above the ground state level and is expressed as
$$
|n,b> = \sum_{\rm distinct} a_{i_1}^\dg \cdots a_{i_n}^\dg |0,b>
\eqn\hb$$
a form identical to the fermionic one. This is a property specific to
the fully antisymmetric (one-hole) state.

In the interacting system we can define raising and lowering
operators 
$$
a_i = {1 \over \sqrt 2} \Bigl(x_i + i p_i -\sum_{j\neq i}
{l\over x_i - x_j} M_{ij} \Bigr)~,~~~
a_i^\dg = {1 \over \sqrt 2} \Bigl(x_i - i p_i +\sum_{j\neq i}
{l\over x_i - x_j} M_{ij} \Bigr)
\eqn\a$$
where $M_{ij}$ is the operator exchanging particles $i$ and $j$.
These satisfy
$$
[ a_i , a_j ] = [a_i^\dg , a_j^\dg ] = 0~,~~~
[a_i , a_j^\dg ] = \delta_{ij} (1+l\sum_{k\neq i} M_{ik})
+ (1-\delta_{ij} )l M_{ij}
\eqn\ca$$
The hamiltonian can be written as
$$
H = \sum_{i=1}^N a_i^\dg a_i
\eqn\ha$$
and upon acting on symmetric states, on which $M_{ij} =1$, it coincides
with \H\ up to a constant equal to the ground state energy. The ground
state
$$
\psi_o = \prod_{i>j} | x_i - x_j |^l e^{-\half \sum_i x_i^2}
\eqn\gr$$
is annihilated by all lowering operators, and all excited states
can be obtained by acting with symmetric combinations of raising
operators. The corresponding (quasi)particle and (quasi)hole states,
thus, can be obtained by acting with the corresponding {\it bosonic}
combination. In particular, the fermionic states can be obtained by
acting with the bosonic combinations of the above operators with
$l=1$. This achieves, therefore, a `bosonization' of the system.
(Note that the usual lowering operators do {\it not} individually 
annihilate the vacuum in the fermionic case, while the above ones 
still do.)

The one-hole states thus are given by the expression \hb\ in terms
of the generalized operators. The resulting wavefunction is a polynomial
in the coordinates $h(x_i )$ multiplying the ground state. To streamline
its derivation, we introduce the commuting operators
$$
A_i = e^{-{l\over 2}a^2} a_i^\dg ~e^{{l\over 2}a^2} = a_i^\dg -
l a~,~~{\rm where}~~~a = \sum_{i=1}^N a_i
\eqn\A$$
it terms of which the hole state can be written
$$
|n,l> = \chi_n (a_i^\dg) |0,l> = e^{{l\over 2} a^2} \chi_n (A_i ) |0,l>
\eqn\hA$$
Each $A_i$ ($= -a_i +\sqrt2 x_i -l a$) acting on the vacuum gives 
$\sqrt2 x_i$, and satisfies
$$
[x_i , A_j ] = {l\over \sqrt2} (1-M_{ij} )~,~~~i \neq j
\eqn\xA$$
Since in $\chi_n (A_i )$ each index $i$ appears at most once in each
term, the commutator \xA\ which arises upon reordering such terms
acting on the vacuum commutes with all remaining operators. Therefore,
it can be pulled through to act on the vacuum, giving zero. The net 
result is that $\chi_n (A_i )|0,l>=2^{n/2} \chi_n (x_i ) |0,l>$. The
operator $a$, on the other hand, acts as a derivative on each $x_i$,
that is
$$
[a, \chi_n (x_i ) ] = {1\over \sqrt2} \sum_j {\partial \over \partial x_j}
\chi_n (x_i ) = {1\over \sqrt2} (N-n+1) \chi_{n-1} (x_i )
\eqn\achi$$
The final result for the polynomial part of the wavefunction is
$$
h_n (x_i ) = 2^{n \over 2} \sum_{k=0}^\infty \left(
{l \over 4}\right)^k {(N-n+2k)! \over k! (N-n)!} \sum_{\rm distinct} 
x_{i_1} \cdots x_{i_{n-2k}}
\eqn\wf$$
(we assume that for negative index the characters vanish).

The use of raising and lowering operators in the derivation of the
wavefunctions was of mainly conceptual advantage. Indeed, these states
could have been derived from Schr\"odinger's equation, starting with
$\chi_n (x_i )$ and recursively generating the other terms. The
advantage of this formalism becomes much more substantial, however,
when calculating the norms of these states. To do that, consider the
operators
$$
a_i (s) = a_i + s
\eqn\as$$
satisfying the same commutation relations as the $a_i$. All hole states
are generated from the function of $s$
$$
|Z(s)> = \prod_{i=1}^N a_i^\dg (s) |0,l>
\eqn\Zs$$
and it suffices to calculate the norm of $|Z(s)>$. To this end, define
the matrix elements
$$
Z_n = <0,l| a_{i_n} (s) \cdots a_{i_1} a_{i_1}^\dg (s) \cdots 
a_{i_n}^\dg (s) |0,l>
$$
$$
Y_n = <0,l| a_{i_n} (s) \cdots a_{i_2} a_{i_1}^\dg (s) \cdots 
a_{i_n}^\dg (s) |0,l>
\eqn\ZY$$
Clearly the above elements are independent of the specific choice
of (distinct) indices and depend only on $s$. By commuting through the 
operators $a_{i_1} (s)$, $a_{i_1}^\dg (s)$ in $Z_n$, $Y_n$ and using
the vacuum condition $a_i (s) |0,l> = s |0,l>$, we obtain the recursion
relations
$$
Z_n = \left[ 1+l(N-n) \right] Z_{n-1} + s Y_n
$$
$$
Y_n = s^* Z_{n-1} -l(n-1) Y_{n-1}
\eqn\recur$$
Solving the above relations with initial conditions $Z_0 =1$, $Y_0 =0$,
we obtain
$$
<Z(s)|Z(s)> = Z_N = \sum_{n=0}^N (s s^* )^n {N \choose n} 
\prod_{k=0}^{N-n-1} (1+lk)
\eqn\Znorm$$
{}From \Znorm\ we can simply read off the norms of the hole states
$$
< h_n | h_n > = {N! \over n! (N-n)!} \prod_{k=0}^{n-1} (1+lk)
\eqn\hnorm$$
In the large-$N$ limit where $N-n \gg 1$ ($n$ need not be small),
the limiting form of the above norms is
$$
< h_n | h_n > = {N! \over \Gamma ({1\over l}) (N-n)!} l^n n^{1-l\over l}
\eqn\largeN$$

It is convenient to express the above set of states in terms of a
generating function. Define the differently normalized characters
$$
\omega_n (x_i ) = (N-n)! \sum_{\rm distinct} x_{i_1}
\cdots x_{i_n}
\eqn\om$$
Then the generating function for the hole states $h_n$ is
$$
h(s) = \sum_{n=0}^N s^n 2^{-{n\over 2}} (N-n)! h_n = e^{{l \over 4} s^2}
\omega (s) = e^{{l \over 4} s^2} \sum_{n=0}^N s^n \omega_n
\eqn\gen$$
(It is understood that only the first $N$ powers of $h(s)$ are
actually energy eigenfunctions.)

Since each $\omega_n$ is homogeneous in $x_i$ with degree $n$, a
rescaling of the $x_i$ amounts to a rescaling of $s$ and thus to
a rescaling of $l$. We conclude from \gen
$$
h_n (x_i;l) = l^{n\over 2} h_n({x_i \over \sqrt l};1)
\eqn\resc$$
Thus the polynomial part of the hole wavefunction is simply a
rescaling of the fermion hole wavefunction.

The hamiltonian of the Calogero model can be separated into center
of mass and relative coordinates. The above hole states are, in general,
not eigenstates of the center of mass motion, but rather superpositions
of center of mass oscillations of energies from 0 to $n$. To isolate
the center of mass coordinate $x$, consider the center-of-mass frame
coordinates $y_i$
$$
y_i = x_i -x = x_i - {1 \over N} \sum_j x_j
\eqn\yi$$
Using \achi\ we can expand
$$
\omega_n (x_i ) = \omega_n (y_i + x) = \sum_k {x^k \over k!}
\omega_{n-k} (y_i )
\eqn\expand$$
The generating function $h(s)$ then becomes
$$
h(s) = e^{{l \over 4} s^2 + sx} \sum_{n=0}^\infty s^n \omega_n (y_i )
\eqn\newh$$
Finally, using the generating function for the Hermite polynomials
$$
\sum_{n=0}^\infty {s^n \over 2^{n \over 2} n!} H_n (x) = e^{sx -{s^2 \over 4}}
\eqn\genH$$
we obtain
$$
h(s) = e^{{1\over 4}(l+{1\over N}) s^2} \omega(s;y_i) \sum_n \left(
{s \over \sqrt2 N}\right)^n H_n ({\sqrt N} x)
\eqn\hH$$
$H_n({\sqrt N}x)$ is an eigenstate of the center of mass oscillation
(the frequency being the same, but the mass being $\sqrt N$). The
remaining part, being a function only of relative coordinates,
has no center of mass excitations, and therefore the above is a generating
function for the energy eigenstates {\it separately} for each $H_n$.
So we obtain 
$$
{\bar h} (s) = \sum_n s^n {\bar h}_n (x_i ) =
e^{{1\over 4}(l+{1\over N}) s^2} \sum_n s^n \omega_n (x_i -x)
\eqn\bareh$$
where ${\bar h}_n (x_i )$ are the `bare' hole wavefunctions, stripped
of all center of mass excitations. We stress that the original hole
states $h_n$ are superpositions of states of the form ${\bar h}_k H_{n-k}$.
Note also that ${\bar h}_1 =0$, since the relative coordinates sum to zero.
It is interesting that moving to the center of mass coordinate essentially
amounts to a shift in the coupling constant $l$ by $1/N$.

Concluding, we remark that the above techniques could be generalized further
to obtain more general classes of states for the model. Such results 
for the Calogero model are encouraging and suggest that a treatment of
the properties of the inverse-square system in the thermodynamic limit
in the operator formalism may be feasible. Other open questions, such
as the existence of a duality symmetry between the Sutherland models
with couplings $l$ and $l^{-1}$ [\G,\MPA,\LPS], of which we have no 
realization yet in the Calogero model, are issues for further investigation.

\ack{I would like to thank T.H.\ Hansson, J.M.\ Leinaas and H.\ Kjonsberg 
for discussions.}

\refout
\end